\documentclass[fleqn,11pt]{article}
\usepackage{epsfig}
\usepackage{floatrow}
\usepackage{xcolor}
\input{amssym.def}
\input{amssym.tex}
\author{ }
\date{ }
\makeatletter

\def\gsim{\compoundrel>\over\sim}
\def\lsim{\compoundrel<\over\sim}
\def\compoundrel#1\over#2{\mathpalette\compoundreL{{#1}\over{#2}}}
\def\compoundreL#1#2{\compoundREL#1#2}
\def\compoundREL#1#2\over#3{\mathrel
      {\vcenter{\hbox{$\m@th\buildrel{#1#2}\over{#1#3}$}}}}
\begin{document}
\vspace*{0.5cm}

\hfill\ {\Large\bf KEK Preprint 2018-4}

\vspace*{3mm}
\hfill\ {\Large\bf May 2018~~~~~~~~~~~~~~~\,}

\vspace*{3mm}
\hfill\ {\Large\bf H~~~~~~~~~~~~~~~~~~~~~~~~~~}

\vspace*{2.0cm}

\vspace*{2.0cm}
\begin{center}
{\Huge\bf Versatile~$e^{+}e^{-}/\gamma\gamma /ep$~facilities}\\
\vspace*{0.6cm}
{\Huge\bf at~a~Future~Circular~Collider}\\
\vspace*{4.4cm}
{\Large R. BELUSEVIC}\\
\vspace*{0.8cm}
{\large {\em High Energy Accelerator Research Organization} (KEK)}\\
\vspace*{1.3mm}
{\large 1-1 {\em Oho, Tsukuba, Ibaraki} 305-0801, {\em Japan}} \\
\vspace*{1.3mm}
{\large r.belusevic@gmail.com}\\
\vspace*{1.3mm}
{\large https://www.rbelusevic-webpage.net}
\end{center}

\thispagestyle{empty}

\newpage

\tableofcontents
\addtocontents{toc}{\protect\vspace{1.3cm}}
\vspace*{5mm}
\noindent
{\large\bf References}


\newpage

\vspace*{0.5cm}
\begin{center}
\begin{minipage}[t]{13.6cm}
{\bf Abstract\,:}\hspace*{3mm}
{This note describes two versatile accelerator complexes that could be
built at a Future Circular Collider (FCC) in order to produce $e^{+}e^{-}$,
$\gamma\gamma$ and $ep$ collisions. The first facility is an SLC-type machine
comprising a superconducting L-band linear accelerator (linac) and two arcs of
bending magnets inside the FCC tunnel. Accelerated by the linac, electron and
positron beams would traverse the arcs in opposite directions and collide at
centre-of-mass energies considerably exceeding those attainable at circular
$e^{+}e^{-}$ colliders. The proposed SLC-type facility would have the same
luminosity as a conventional two-linac $e^{+}e^{-}$ collider. The L-band linac may
form a part of the injector chain for a 100-TeV proton collider inside the FCC
tunnel (FCC-pp), and could deliver electron or positron beams for an $ep$ collider
(FCC-ep). The second facility is an ILC-based $e^{+}e^{-}$ collider placed
tangentially to the circular FCC tunnel. If the collider is positioned
asymmetrically with respect to the FCC tunnel, electron (or positron) bunches
could be accelerated by both linacs before they are brought into collision with
the 50-TeV beams from the FCC-pp proton storage ring. The two linacs may also form
a part of the injector chain for FCC-pp. Each facility could be converted into a
$\gamma\gamma$ collider or a source of multi-MW beams for fixed-target experiments.}
\end{minipage}
\end{center}

\vspace*{0.5cm}
\renewcommand{\thesection}{\arabic{section}}
\section{Introduction}
\vspace*{0.3cm}

\setcounter{equation}{0}

~~~~The maximum luminosity at a circular $e^{+}e^{-}$ collider, such as the
proposed FCC-ee facility \cite{blondel}, is severely constrained by beamstrahlung
effects at high energies; also, it is very difficult to achieve a high degree of
beam polarization \cite{koratzinos}. At the $e^{+}e^{-}$ facilities described in
this paper, luminosity grows almost linearly with the beam energy \cite{boscolo}
and the initial electron beam polarization can reach about 80\% \cite{aurand}. The
availability of polarized beams is essential for some important precision
measurements in $e^{+}e^{-}$ and $\gamma\gamma$ collisions \cite{moortgat}.

The rich set of final states in $e^{+}e^{-}$ and $\gamma\gamma$ collisions would
play an essential role in measuring the mass, spin, parity, two-photon width and
trilinear self-coupling of the {\em Standard Model} (SM) Higgs boson, as well as
its couplings to fermions and gauge bosons. Some of those measurements require
centre-of-mass (c.m.) energies $\sqrt{s_{ee}}$ considerably exceeding those
attainable at circular $e^{+}e^{-}$ colliders. For instance, one has to measure
separately the HWW, HHH and Htt couplings at $\sqrt{s_{ee}} \gsim 500$ GeV in
order to determine the corresponding SM loop contributions to the effective HZZ
coupling \cite{McCullough}. This would not be possible to accomplish using the
proposed FCC-ee facility.

\begin{figure}[t]
\begin{center}
\epsfig{file=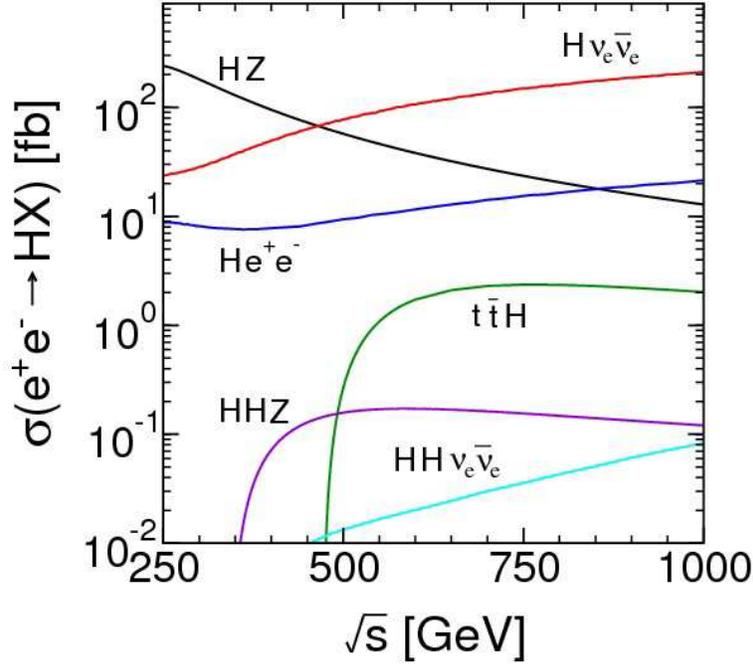,width=0.62\textwidth}
\end{center}
\vskip -7mm
\caption{Centre-of-mass energy dependence of various cross-sections for single and
double SM Higgs-boson production in $e^{+}e^{-}$ annihilations
\cite{asner0}.}
\label{fig:Xsections3}
\end{figure}

The Htt coupling cannot be {\em directly measured} in $e^{+}e^{-}$ interactions
below $\sqrt{s_{ee}} \approx 500$ GeV, since the cross-section for the relevant
process is negligible (see Fig.\,\ref{fig:Xsections3}). Concerning the HHH
coupling, this quantity can be {\em directly measured} at energies above the
kinematic threshold for the reaction $e^{+}e^{-} \rightarrow {\rm ZHH}$, or by
using the WW-fusion channel at $\sqrt{s_{ee}} \gsim 1$ TeV. {\em Indirect} and
{\em model dependent} measurements of the HHH coupling are possible at lower
energies by exploiting the loop corrections to single Higgs channels. However,
the sensitivity of such measurements is very low, as can be inferred from Fig.\,4
in \cite{Di Vita}.

Since the Higgs-boson mass affects the values of electroweak observables through
radiative corrections, high-precision electroweak measurements provide a natural
complement to direct studies of the Higgs sector. All the measurements made at LEP
and SLC could be repeated at the facilities described in this note, but at much
higher luminosities and using 80\% polarized electron beams \cite{erler}. The
importance of beam polarization for some high-precision measurements was already
stressed.

If electron or positron bunches are brought into collision with the 50-TeV proton
beams from the FCC-pp storage ring, one would obtain an important source of
deep-inelastic $ep$ interactions.\footnote{The proposed FCC-eh electron-proton
collider \cite{bruning} would provide a higher luminosity than the facilities
described in this paper, but would have a considerably lower electron beam energy
(around 60 GeV).} Such interactions would yield valuable information on the
quark-gluon content of the proton, which is crucial for precision measurements at
the FCC-pp. The physics potential of a TeV-scale $ep$ collider is comprehensively
discussed in \cite{LH_ep}.

An SLC-type facility or a conventional two-linac collider could be constructed in
several stages, each with distinct physics objectives that require particular
centre-of-mass energies (see Fig.\,\ref{fig:Xsections3}):

\vspace*{3mm}
\[
       \begin{array}{ll}
\bullet~~e^{+}e^{-} \rightarrow {\rm Z,\,WW};\hspace*{0.5cm}\gamma\gamma 
\rightarrow {\rm H}~~~~~~~&~~~~~~~\sqrt{s_{ee}} \sim 90~{\rm to}~180~{\rm GeV} 
\\*[4mm]
\bullet~~e^{+}e^{-} \rightarrow {\rm HZ}~~~~~~~&~~~~~~~\sqrt{s_{ee}} \sim 250~
{\rm GeV} \\*[4mm]
\bullet~~e^{+}e^{-} \rightarrow t\bar{t};\hspace*{0.5cm}\gamma\gamma \rightarrow
{\rm HH}~~~~~~~&~~~~~~~\sqrt{s_{ee}} \sim 350~{\rm GeV} \\*[4mm]
\bullet~~e^{+}e^{-}\rightarrow {\rm HHZ},\,{\rm H}t\bar{t},\,{\rm H}\nu\bar{\nu}
~~~~~~~&~~~~~~~\sqrt{s_{ee}} \gsim 500~{\rm GeV}
       \end{array} \]

For some processes within and beyond the SM, the required c.m. energy is
considerably lower in $\gamma\gamma$ collisions than in $e^{+}e^{-}$ or
proton-proton interactions. For example, the heavy neutral MSSM Higgs bosons can
be created in $e^{+}e^{-}$ annihilations only by associated production
($e^{+}e^{-} \rightarrow H^{0}A^{0}$),  whereas in $\gamma\gamma$ collisions they
are produced as single resonances ($\gamma\gamma \rightarrow H^{0},\,A^{0}$) with
masses up to 80\% of the initial $e^{-}e^{-}$ collider energy.

It is straightforward to convert an SLC-type facility or a conventional two-linac
collider into a high-luminosity $\gamma\gamma$ collider with highly polarized
beams. The CP properties of any neutral Higgs boson produced at a photon
collider can be directly determined by controlling the polarizations of
Compton-scattered photons (see \cite{belusevic} and references therein).

\vspace*{0.3cm}
\section{An SLC-type facility at FCC for creating $e^{+}e^{-}/\gamma\gamma /ep$ collisions}
\vspace*{0.3cm}

\begin{figure}[!h]
\begin{center}
\epsfig{file=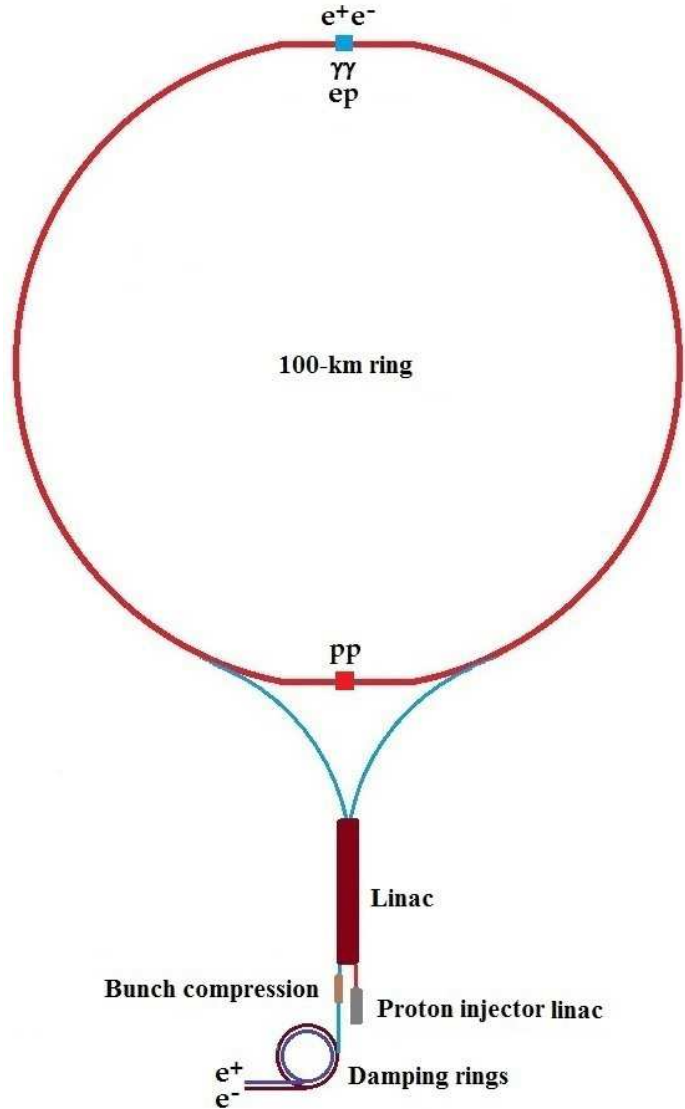,width=0.56\textwidth}
\end{center}
\vskip -5mm
\caption{Schematic layout of an SLC-type facility at FCC \cite{belusevic}. The
superconducting L-band linac could be a part of the FCC-pp injector chain
(how to adjust the proton bunch length throughout the accelerator complex is
discussed in Section 4). The entire facility would serve as a source of
$e^{+}e^{-}$, $\gamma\gamma$, $pp$ and $ep$ interactions.}
\label{fig:SLC}
\end{figure}

~~~~A schematic layout of an SLC-type $e^{+}e^{-}/\gamma\gamma$ facility at
a Future Circular Collider (FCC) is shown in Fig.\,\ref{fig:SLC}. Damped and
bunch-compressed electron and positron beams, accelerated by a single
superconducting L-band linac, traverse two arcs of bending magnets in opposite
directions and collide at an interaction point surrounded by a detector. The
beams are then disposed of, and this machine cycle is repeated at a rate of up to
10 Hz. In contrast to a conventional two-linac collider, an SLC-type machine would
have a single bunch compression system.

The linac in Fig.\,\ref{fig:SLC} would contain ILC-type superconducting L-band
cavities placed within cryogenic vessels and fed by multi-beam klystrons. The
current design for the {\em International Linear Collider} (ILC), based on the
technology originally developed at DESY, uses L-band (1.3 GHz) superconducting
niobium rf cavities that have average accelerating gradients of 31.5 MeV/m (see
\cite{ILC} and references therein). Nine cavities, each 1 m long, are mounted
together in a string and assembled into a common low-temperature cryostat or
{\em cryomodule}. Liquid helium is used to cool cavities to $-271^{\circ}$ C.

An ILC-type main linac is composed of rf units, each of which is formed by three 
contiguous cryomodules containing 26 nine-cell cavities. Every unit has an rf
source, which includes a pulse modulator, a 10 MW multi-beam klystron, and a
waveguide system that distributes the power to the cavities. An ILC-type design
has the following characteristics:

\vspace*{2mm}
$\bullet$~~Transverse wakefield effects are drastically reduced due to the large
size of L-band cavities, which means that cavity alignment tolerances can be
relaxed. This is particularly relevant for an SLC-type facility, where both
$e^{+}$ and $e^{-}$ bunches are alternately accelerated;

\vspace*{1mm}
$\bullet$~~Superconducting ILC-type rf cavities can be loaded using a long rf
pulse ($\sim 1.5$ ms) from a source with low peak rf power;

\vspace*{1mm}
$\bullet$~~Wall-plug to beam power transfer efficiency is about twice that of
X-band cavities, for example, which makes an ILC-type linac much cheaper to
operate;

\vspace*{1mm}
$\bullet$~~The long rf pulse allows a long `bunch train' ($\sim 1$ ms), with many
bunches ($\sim 3000$) and a relatively large bunch spacing ($\sim 300$ ns). A
trajectory correction (feedback) system within the train can therefore be used
to bring the beams into collision.
\vspace*{2mm}

At an SLC-type facility, $e^{+}$ and $e^{-}$ bunches are alternately accelerated
inside the linac. In order to focus $e^{-}$ and $e^{+}$ beams in the two
transverse planes, quadrupole magnets with alternating polarities have to be
placed along the linac. The $x$-plane for electrons is then like the $y$-plane
for positrons, and vice versa. At the {\em Stanford Linear Collider} (SLC),
emittance growth in the linac was reduced using the beam diagnostics obtained by
steering both beams together \cite{zimmermann}.

The spacing between electron bunches in a superconducting L-band accelerator can
be made to match that between proton bunches in the FCC-pp storage ring. Also,
the length of an electron `bunch train' corresponds roughly to the FCC ring circumference.
An ILC-type linac is thus a suitable source of electron beams for
a high-luminosity $ep$ collider. This is not the case with an X-band linac, where
the electron bunch spacing ($\sim 1$ ns) is much shorter than that between proton
bunches at the FCC-pp (see Table 2).

The total energy loss due to {\em synchrotron radiation} in each arc of the
proposed SLC-type facility is $\Delta{\rm E} = 14.4$ GeV for ${\rm E}_{0} = 250$
GeV and $\rho = 12$ km, where ${\rm E}_{0}$ is the initial electron (or positron)
beam energy and $\rho$ the {\em effective bending radius}. Synchrotron radiation is
also responsible for {\em energy spread}, $\mbox{\large{$\sigma$}}_{\rm E}^{~}/
{\rm E}$, in an electron beam. A plot of $\mbox{\large{$\sigma$}}_{\rm E}^{~}/
{\rm E}$ as a function of beam energy for $\rho = 12$ is shown in Fig.\,6 of \cite{belusevic}.
For electron beams with ${\rm E} \lsim 450$ GeV traversing the
bending arcs of the proposed SLC-type facility, the increase in the
{\em horizontal beam emittance} would not exceed $2\,\mu m$, the value of this
parameter at the KEK--ATF damping ring (see Fig.\,7 in \cite{belusevic}). Figures
6 and 7 in \cite{belusevic} were produced using a lattice of combined-function
FODO cells as an input to K. Oide's SAD tracking code.

The {\em geometric luminosity} at a conventional two-linac collider is given by
\begin{equation}
{\cal L}_{ee} \,\propto\, \frac{\gamma N_{e}^{\,2}N_{b}f_{\rm rep}^{~}}{\sqrt
{(\mbox{\large{$\varepsilon$}}_{x}^{n}\beta_{x}^{*})(\mbox{\large{$\varepsilon$}}
_{y}^{n}\beta_{y}^{*})}} \,\equiv\, \frac{{\cal P}_{b}^{~}}{\sqrt{s_{ee}}}
\frac {\gamma N_{e}^{~}}{\sqrt{(\mbox{\large{$\varepsilon$}}_{x}^{n}\beta_{x}^{*})
(\mbox{\large{$\varepsilon$}}_{y}^{n}\beta_{y}^{*})}}
\end{equation}
where $\beta_{x}^{*}$ and $\beta_{y}^{*}$ are the horizontal and vertical
{\em beta functions} at the interaction point (IP), respectively, $\mbox{\large
{$\varepsilon$}}_{x}^{n}$ and $\mbox{\large{$\varepsilon$}}_{y}^{n}$ are the
normalized transverse {\em beam emittances}, $N_{e}$ is the number of electrons in
a 'bunch', $N_{b}$ is the number of bunches per rf pulse, $f_{\rm rep}^{~}$ is the
pulse {\em repetition rate}, $\sqrt{s_{ee}}$ is the c.m. energy, ${\cal P}_{b}^{~}
= N_{e}N_{b}f_{\rm rep}^{~}\sqrt{s_{ee}}$ is the {\em beam power}, and $\gamma
\equiv {\rm E}/m_{e}c^{2}$ is the {\em Lorentz factor} of the electron beam with
energy E.

There are $N_{b}/2$ electron or positron bunches in each arc of an SLC-type
facility. If its repetition rate is twice that of a conventional two-linac
collider, so that the same wall-plug power is used, the two machines would have
the same luminosity (see Eq. (1)).

An important feature of the proposed facilities is the possibility of using backscattered
laser beams to produce high-energy $\gamma\gamma$ collisions
\cite{ginzburg}. In order to attain maximum luminosity at a $\gamma\gamma$
collider, every electron bunch in the accelerator should collide with a laser
pulse of sufficient intensity for $63\%$ of the electrons to undergo a Compton
scattering. This requires a laser system with high average power, capable of
producing pulses that would match the temporal spacing of electron bunches. These
requirements could be satisfied by an optical {\em free electron laser}
\cite{saldin}. The luminosity of a gamma-gamma collider is ${\cal L}_
{\gamma\gamma} = (N_{\gamma}^{~}/N_{e}^{~})^{2\,}{\cal L}_{ee} \approx (0.63)^
{2\,}{\cal L}_{ee}$, where $N_{\gamma}^{~}$ is the number of backscattered laser
photons.

\vspace*{0.3cm}
\section{An ILC-based $e^{+}e^{-}/\gamma\gamma /ep$ facility at FCC}
\vspace*{0.3cm}

~~~~The ILC-based facility at a Future Circular Collider (FCC) shown in
Fig.\,\ref{fig:2linac} features a superconducting two-linac $e^{+}e^{-}$ collider
placed tangentially to the FCC tunnel  \cite{belusevic2}. Using an optical free-electron
laser, the linacs could be converted into a high-luminosity $\gamma\gamma$ collider.

As mentioned in the Introduction, the maximum luminosity at a circular
$e^{+}e^{-}$ collider is severely constrained by beamstrahlung effects at high
energies; also, it is very difficult to achieve a high degree of beam
polarization. At the $e^{+}e^{-}$ facilities described in this paper, luminosity
grows almost linearly with the beam energy and the electron beam polarization can
reach 80\%.
 
The maximum c.m. energy of an SLC-type facility is limited by the radius of the
bending arcs, which is not the case with a conventional two-linac $e^{+}e^{-}$
collider. Note also that the asymmetric accelerator configuration in
Fig.\,\ref{fig:2linac} allows one to double the energy of electron or positron beams
before their extraction (see Section 4).

The baseline parameters for the proposed ILC collider, shown in Table 1, reflect
the need to balance the constraints imposed by the various accelerator
sub-systems, as explained in \cite{ILC-TDR}. The rf power is provided by 10 MW
multi-beam klystrons, each driven by a 120 kV pulse modulator. The estimated AC
power is 122 MW at $\sqrt{s_{ee}}=250$ GeV and 163 MW at $\sqrt{s_{ee}}=500$ GeV.

In order to maximize luminosity at low centre-of-mass energies, the beam power
could be increased by increasing the pulse repetition rate $f_{\rm rep}^{~}$ 
while reducing the accelerating gradient of the main linacs. At $\sqrt{s_{ee}} =
250$ GeV, the power consumption of the main 250-GeV linacs is reduced by over a
factor of two when they are running at half their nominal gradient. Under these
conditions, one can run the accelerator at the maximum repetition rate of 10 Hz
(determined by the cryogenic system and the beam damping time  $t_{\rm damp}
\approx 80$ ms), thus doubling its luminosity.

\begin{table}[!t]
\centering
\ttabbox{\caption{Baseline ILC parameters \cite{ILC-TDR}}\label{tab:table1}}
   {\begin{tabular}{lllrr}
\noalign{\vskip 1mm}   
\hline
\hline
\noalign{\vskip 1mm}
Centre-of-mass energy & $\sqrt{s_{ee}}$ & GeV & 250 & 500\\
\noalign{\vskip 1mm}
\hline
\noalign{\vskip 1mm}
Pulse repetition rate & $f_{\rm rep}^{~}$ & Hz & 5 & 5\\
\noalign{\vskip 1mm}
\hline
\noalign{\vskip 1mm}
Bunch population & $N_{e}$ & $\times 10^{10}$ & 2 & 2\\
\noalign{\vskip 1mm}
\hline
\noalign{\vskip 1mm}
Number of bunches & $N_{b,e}$ & ~ & 1312 & 1312\\
\noalign{\vskip 1mm}
\hline
\noalign{\vskip 1mm}
Bunch interval & $\Delta t_{b,e}^{~}$ & ns & 554 & 554\\
\noalign{\vskip 1mm}
\hline
\noalign{\vskip 1mm}
RMS bunch length & $\sigma_{z,e}^{~}$ & mm & 0.3 & 0.3\\
\noalign{\vskip 1mm}
\hline
\noalign{\vskip 1mm}
Norm. horizontal emittance at IP & $\mbox{\large{$\varepsilon$}}_{x}^{n}$ &
$\mu$m & 10 & 10\\
\noalign{\vskip 1mm}
\hline
\noalign{\vskip 1mm}
Norm. vertical emittance at IP & $\mbox{\large{$\varepsilon$}}_{y}^{n}$ &
nm & 35 & 35\\
\noalign{\vskip 1mm}
\hline
\noalign{\vskip 1mm}
Horizontal beta function at IP & $\beta^{*}_{x}$ & mm & 13 & 11\\
\noalign{\vskip 1mm}
\hline
\noalign{\vskip 1mm}
Vertical beta function at IP & $\beta^{*}_{y}$ & mm & 0.41 & 0.48\\
\noalign{\vskip 1mm}
\hline
\noalign{\vskip 1mm}
RMS horizontal beam size at IP & $\sigma_{x}^{*}$ & nm & 729 & 474\\
\noalign{\vskip 1mm}
\hline
\noalign{\vskip 1mm}
RMS vertical beam size at IP & $\sigma_{y}^{*}$ & nm & 7.7 & 5.9\\
\noalign{\vskip 1mm}
\hline
\noalign{\vskip 1mm}
Vertical disruption parameter & $D_{e}$ & ~ & 24.5 & 24.6 \\
\noalign{\vskip 1mm}
\hline
\noalign{\vskip 1mm}
Luminosity & ${\cal L}_{ee}$ & $\times 10^{34}~cm^{-2}\,s^{-1}$ & 0.75 & 1.8\\
\noalign{\vskip 1mm}
\hline
\hline
   \end{tabular}}
\end{table}

The two superconducting L-band linacs in Fig.\,\ref{fig:2linac} may also form a
part of the FCC-pp injector chain. Since the collider is positioned
asymmetrically with respect to the FCC tunnel, electron (or positron) bunches
could be accelerated by both linacs before they are brought into collision with
the 50-TeV beams from the FCC-pp proton storage ring, as mentioned above. The
entire accelerator complex would serve as a source of $e^{+}e^{-}$,
$\gamma\gamma$, $pp$ and $ep$ interactions.

\vspace*{0.3cm}
\section{Main parameters of a linac-ring $ep$ collider at FCC}
\vspace*{0.3cm}

~~~~The idea to combine a 140-GeV electron linac and a 20-TeV proton storage ring
in order to produce $ep$ interactions at very high c.m. energies was put forward
in 1979 as a possible option at the SSC proton collider \cite{weber}. In 1987 it
was proposed to place a 2-TeV linear $e^{+}e^{-}$ collider (VLEPP) tangentially
to a 6-TeV proton-proton collider (UNK) at IHEP in Protvino \cite{alekhin}, with
the aim of obtaining both $ep$ and $\gamma p$ collisions. Similar proposals for
lepton-hadron and photon-hadron colliders at HERA, LHC and FCC have since been
made (see \cite{akay} and references therein).

The SLC-type facility in Fig.\,\ref{fig:SLC} could be used to produce TeV-scale
$ep$ interactions. Accelerated by the superconducting L-band linac, electron or
positron beams would traverse one of the two arcs of bending magnets inside the
FCC tunnel and collide with the FCC-pp proton beams.

\begin{figure}[!t]
\begin{center}
\epsfig{file=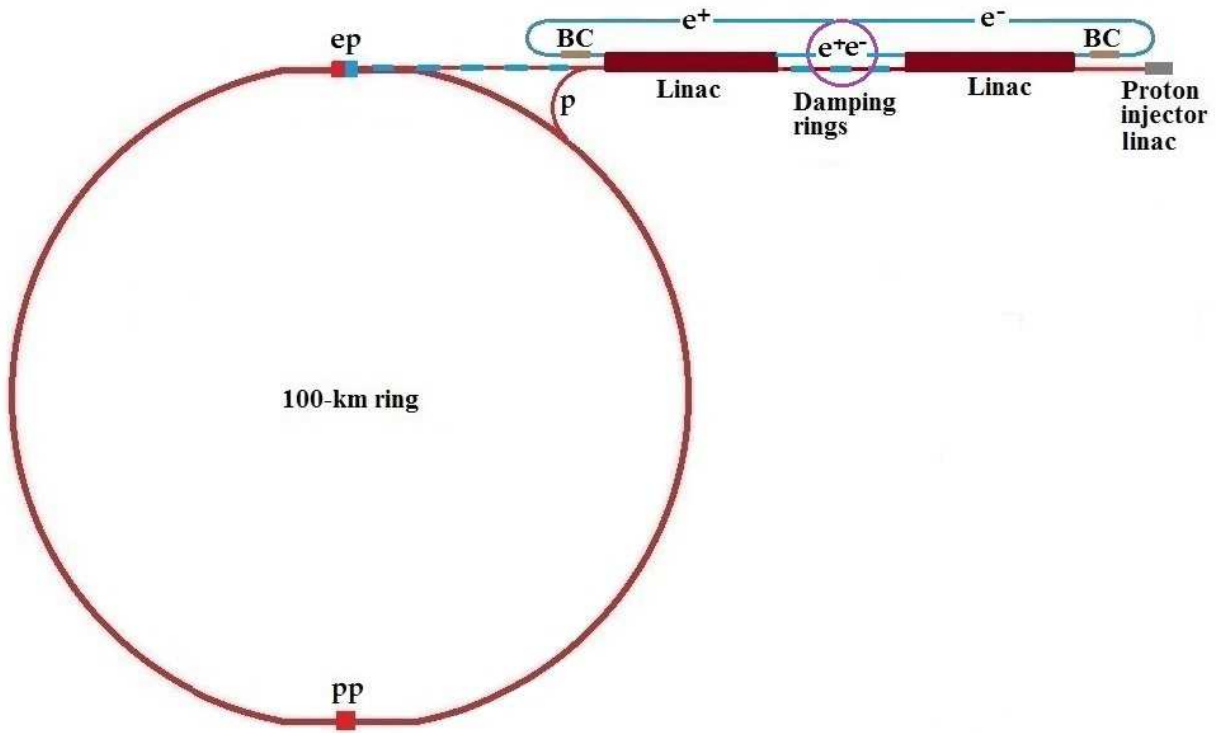,width=0.99\textwidth}
\end{center}
\vskip -5mm
\caption{An ILC-based facility at FCC (BC stands for {\em bunch compression}).
Electron (or positron) bunches are accelerated by both linacs before their
collision with the 50-TeV proton beam from the FCC-pp storage ring. The two
superconducting L-band linacs may form the low-energy part of the FCC-pp injector
chain \cite{belusevic2}. The issue of proton bunch length is discussed in Section 4.}
\label{fig:2linac}
\end{figure}

The facility shown in Fig.\,\ref{fig:2linac} is an ILC-based version of the
original VLEPP$\otimes$UNK design. Since the collider is positioned asymmetrically
with respect to the FCC tunnel, electron (or positron) bunches could be
accelerated by both linacs (which contain {\em standing wave cavities}) before
they are brought into collision with the 50-TeV beams from the FCC-pp proton
storage ring.

In Section 2 it was noted that an ILC-type linac is a suitable source of
electron beams for an electron-proton collider, because: (1) the spacing between
electron bunches can be made to match that between the proton bunches in the
FCC-pp storage ring, and (2) the length of an electron `bunch train' corresponds
roughly to the FCC ring circumference.

In {\em head-on collisions} of ultra-relativistic electrons and protons, the
centre-of-mass energy is $\sqrt{s_{ep}} = 2\sqrt{{\rm E}_{e}{\rm E}_{p}}$.
The total electron beam current $I_{e} = {\cal P}_{e}/{\rm E}_{e}$ is limited by
the maximum allowed beam power ${\cal P}_{e}$ for a given electron beam energy
${\rm E}_{e}$. Assuming that round electron and proton beams of equal transverse
sizes are colliding head-on at the interaction point (IP),\footnote{The two beams
are chosen to have roughly equal transverse sizes in order to reduce adverse
effects a much smaller electron beam could have on the proton beam lifetime.
Electron bunches are discarded after each collision.} the luminosity of the
collider is given by
\cite{tigner}\cite{zimmermann1}
\begin{equation}
{\cal L}_{ep} = f_{c}^{~}\frac{N_{e}N_{p}}{4\pi\sigma^{2}_{p}}\,{\cal H} \equiv
\frac{I_{e}}{4\pi{\rm e}}\frac{N_{p}}{\mbox{\large{$\varepsilon$}}^{n}_{p}}
\frac{\gamma_{p}^{~}}{\beta^{*}_{p}}\,{\cal H}
\end{equation}
In these expressions, $N_{e}$ and $N_{p}$ are the electron and proton bunch
populations, respectively; $f_{c}^{~}$ is the bunch collision frequency;
${\cal H}$ is a correction factor discussed below; and $\sigma^{~}_{p} = \sqrt
{\mbox{\large{$\varepsilon$}}^{n}_{p}\beta^{*}_{p}/\gamma_{p}^{~}}$ is the proton
beam size at IP, expressed in terms of the normalized proton beam  emitance,
$\mbox{\large{$\varepsilon$}}^{n}_{p}$, the proton beta  function at IP,
$\beta^{*}_{p}$, and the Lorentz factor of the proton beam, $\gamma_{p}^{~}$.
Note that the luminosity is proportional to the electron beam power
${\cal P}_{e} = {\rm e}N_{e}f_{c}^{~}{\rm E}_{e} = I_{e}{\rm E}_{e}$ (e is the
electron charge), the proton beam energy ($\gamma_{p}^{~}$), and the proton
beam brightness $N_{p}/\mbox{\large{$\varepsilon$}}^{n}_{p}$.

In Eq. (2), ${\cal H}$ is a product of three correction factors with values typically close to unity:
\begin{equation}
{\cal H} \equiv H_{\rm hourglass}^{~}\cdot H_{\rm pinch}^{~}\cdot H_{\rm filling}
^{~}
\end{equation}
The factor $H_{\rm filling}$ takes into account the filling patterns of the
electron and proton beams. If the number of proton bunches $N_{b,p} = 10600$ and
the bunch interval $\Delta t_{b,p}^{~} = 25$ ns (see Table 2), the `length' of the
proton beam is $2.65\times 10^{5}$ ns. This corresponds to 80 km, which means
that only 80\% of the FCC circumference is filled with proton bunches
($H_{\rm filling} = 0.8$). In this particular case 20\% of the electron bunches
would not collide with the proton beam.

The factor $H_{\rm hourglass}$ accounts for a loss of luminosity when the
bunch length is comparable to or larger than $\beta^{*}$. The beta function
$\beta (s) = \beta^{*} + s^{2}/\beta^{*}$ grows parabolically as a function of
distance $s$ from the interaction point, which causes the beam size to increase:
\begin{equation}
\sigma (s) = \sqrt{\beta (s)\cdot\mbox{\large{$\varepsilon$}}} \approx s\sqrt
{\mbox{\large{$\varepsilon$}}/\beta^{*}}
\end{equation}
As the beam size increases, the contribution to the luminosity from regions with
large $\sigma$ decreases ({\em hourglass effect}). For zero crossing angle and
$\sigma_{z,p}^{~} \gg \sigma_{z,e}^{~}$,
\begin{equation}
H_{\rm hourglass}^{~}(x) = \sqrt{\pi}\,x\,{\rm e}^{x^{2}\,}{\rm erfc}(x)
\end{equation}
with
\begin{equation}
x \equiv \frac{2\beta^{*}_{e}}{\sigma_{z,p}^{~}}\frac{\mbox{\large
{$\varepsilon$}}_{e}^{~}/\mbox{\large{$\varepsilon$}}_{p}^{~}}{\sqrt{1 + (\mbox
{\large{$\varepsilon$}}_{e}^{~}/\mbox{\large{$\varepsilon$}}_{p}^{~})^{2}}},
\hspace*{2cm}{\rm erfc}(x) = \frac{\raisebox{-.4ex}{2}}{\sqrt{\pi}}\int_{x}^
{\infty}{\rm e}^{-t^{2}}\,{\rm d}t
\end{equation}
where $\mbox{\large{$\varepsilon$}}_{e}^{~}$ and $\mbox{\large{$\varepsilon$}}_
{p}^{~}$ denote {\em geometric emittances} \cite{LH_ep}\cite{furman} (the
normalized emittance $\mbox{\large{$\varepsilon$}}^{n} = \gamma\mbox{\large
{$\varepsilon$}}$ is invariant under acceleration); erfc($z$) is the
`complementary error function' (defined as the area under the 'tails' of a
Gaussian distribution).

\begin{table}[!t]
\centering
\ttabbox{\caption{Baseline FCC-pp parameters \cite{benedikt}\cite{zimmermann3}.
Numbers inside round brackets represent parameters for 5 ns bunch
spacing.}\label{tab:table2}}
   {\begin{tabular}{lllr}
\noalign{\vskip 1mm}   
\hline
\hline
\noalign{\vskip 1mm}   
Beam energy & E$_{p}$ & TeV & 50\\
\noalign{\vskip 1mm}
\hline
\noalign{\vskip 1mm}
Initial bunch population & $N_{p}$ & $\times 10^{10}$ & 10~~(2)\\
\noalign{\vskip 1mm}
\hline
\noalign{\vskip 1mm}
Number of bunches & $N_{b,p}$ & ~ & 10600~~(53000)\\
\noalign{\vskip 1mm}
\hline
\noalign{\vskip 1mm}
Bunch interval & $\Delta t_{b,p}^{~}$ & ns & 25~~(5)\\
\noalign{\vskip 1mm}
\hline
\noalign{\vskip 1mm}
RMS bunch length & $\sigma_{z,p}^{~}$ & mm & 80\\
\noalign{\vskip 1mm}
\hline
\noalign{\vskip 1mm}
Norm. transverse emittance & $\mbox{\large{$\varepsilon$}}^{n}_{p}$
 & $\mu$m & 2.2~~(0.44)\\
\noalign{\vskip 1mm}
\hline
\noalign{\vskip 1mm}
Beta function at IP & $\beta^{*}_{p}$ & m & 0.3\\
\noalign{\vskip 1mm}
\hline
\noalign{\vskip 1mm}
Beam size at IP & $\sigma^{~}_{p}$ & $\mu$m & 6.8~~(3)\\
\noalign{\vskip 1mm}
\hline
\noalign{\vskip 1mm}
Beam-beam tune shift/IP & $\Delta Q_{p}^{~}$ & ~ & 0.005\\
\noalign{\vskip 1mm}
\hline
\noalign{\vskip 1mm}
Luminosity/IP & ${\cal L}_{ep}$ & $\times 10^{32}~cm^{-2}\,s^{-1}$ & 2.3\\
\noalign{\vskip 1mm}
\hline
\hline
   \end{tabular}} 
\end{table}

The enhancement factor $H_{\rm pinch}$ in Eq. (3) is due to the attractive
{\em beam-beam force}. Since the electron bunch charge is relatively small and the
proton energy is high, the beam-beam force acting on electrons has a much greater
strength than that acting on protons. Consequently, the electron bunch is focused
by the protons during a collision,. This leads to a reduction in the transverse
electron beam size (`pinch effect') and hence to an increase in the luminosity.
The effect can be simulated using the program {\em Guinea-Pig} (see \cite{bruning}
and references therein, as well as Table 3 below).

One can ignore the longitudinal structure of electron bunches because they are
much shorter than proton bunches. In this case the {\em transverse disruption} of
the electron beam during a collision is described by the parameter
\cite{yokoya}\cite{hao}
\begin{equation}
D_{e} = \frac{r_{e}^{~}}{\raisebox{.4ex}{$\gamma_{e}^{~}$}}\frac{N_{p}
\sigma_{z,p}}{\sigma_{p}^{2}}
\end{equation}
where $\gamma_{e}^{~}$ is the Lorentz factor of the electron beam, $r_{e}^{~}
\approx 2.82\times 10^{-15}$~m is the classical radius of the electron, and
$\sigma_{z,p}$ is the proton bunch length. For $\beta^{*}_{p} = 10$~cm, the
disruption parameter can be as large as $D_{e} \approx 20$ in an $ep$ linac-ring
collider \cite{zimmermann1}.

As already mentioned, the luminosity of an $ep$ collider is proportional to the
proton {\em beam brightenss} $N_{p}/\mbox{\large{$\varepsilon$}}^{\mbox{\tiny
{N}}}_{p}$ (see Eq. (2)). Together with a given bunch length and energy spread,
the beam brightness is a measure of the phase-space density. In the low-energy
part of a proton injector, the quantity $N_{p}/\mbox{\large{$\varepsilon$}}
^{n}_{p}$ is limited by space-charge forces that induce a
{\em transverse tune shift}\,\footnote{The `tune' or $Q$ value is defined as the number
of betatron oscillations per revolution in a circular accelerator. The charge and current of
a high-inensity beam in an accelerator create self-fields and image fields that alter the beam
dynamics and influence the single-particle motion as well as coherent oscillations of the beam
as a whole. The effect of space-charge forces is to change $Q$ by an amount
$\Delta Q_{sc}$ (`tune shift') \cite{schindl}.}
\begin{equation}
\Delta Q_{sc} \propto \frac{N_{p}}{\mbox{\large{$\varepsilon$}}^{n}_{p}}\frac{1}
{(v_{p}^{~}/c)^{2}\gamma_{p}^{2}}
\end{equation}
Here $v_{p}^{~}$ is the proton velocity and $c$ is the speed of light in vacuo 
\cite{schindl}\cite{garoby}. In order to reduce the effect of space-charge forces
at low energies and deliver proton bunches a few mm long, the facility in
Fig.\,\ref{fig:2linac} features a single 3-GeV proton injector linac similar to
that currently being built at the {\em European Spallation Source} (ESS)
\cite{danared}.

At high energies, the beam brightness in a storage ring slowly diminishes due to
Coulomb scattering of protons within a bunch ({\em intra-beam scattering})
\cite{piwinski}. In the presence of {\em dispersion} (see footnote 4), the
intra-beam scattering also leads to an increase in emittance. This sets the
ultimate limit on the phase-space density in a proton storage ring. The growth of
a beam of charged particles due to intra-beam scattering is characterized by the
horizontal {\em growth rate} \cite{parzen}
\begin{equation}
\mbox{\large{$\tau$}}^{-1}_{x} \propto \frac{N_{p}}{\mbox{\large{$\varepsilon$}}
^{n}_{x}\mbox{\large{$\varepsilon$}}^{n}_{y}\mbox{\large{$\varepsilon$}}^{n}_{l}}
\end{equation}
where $\mbox{\large{$\varepsilon$}}^{n}_{x,y}$ are the normalized beam emittances,
$\mbox{\large{$\varepsilon$}}^{n}_{l} \equiv \beta\gamma\sigma_{z,p}^{~}\sigma_
{\!\Delta p/p}^{~}$ and $\sigma_{\!\Delta p/p}^{~}$ is the r.m.s. relative
momentum $\Delta p/p$. Note that the growth rate depends linearly on the
normalized phase-space density. In the FCC-pp storage ring synchrotron radiation
damping is expected to be much stronger than the intra-beam scattering, making the
latter effect less of an issue \cite{benedikt}.

The space-charge forces that limit the beam brightness are determined by the
longitudinal charge density and thus by the proton bunch length
$\sigma_{z,p}^{~}$. To attain maximum brightness, $\sigma_{z,p}^{~}$ should be as
large as possible. On the other hand, there is a loss of luminosity when the bunch
length is comparable to or larger than $\beta^{*}$ (this {\em hourglass effect}
was described earlier). Furthermore, the transverse disruption of the electron
beam during an $ep$ collision is proportional to $\sigma_{z,p}^{~}$, as shown in
Eq. 7. While optimizing the bunch length within these constraints, the beam
stability must be preserved (see below).

A particle in one colliding beam experiences a force due to the electromagnetic
interactions with all the particles in the opposing beam. This force depends upon
the displacement of the particle from the equilibrium orbit of the opposing bunch.
For small particle displacements, the beam-beam interaction is nearly linear, and
its strength is characterized by a parameter known as the {\em beam-beam tune
shift} \cite{ruggiero}:
\begin{equation}
\Delta Q_{p}^{~} \equiv \frac{r_{p}^{~}}{4\pi}\frac{N_{e}}{\sigma^{2}_{e}}\frac
{\beta^{*}_{p}}{\gamma_{p}^{~}} \approx \frac{r_{p}^{~}N_{e}}{4\pi\mbox{\large
{$\varepsilon$}}^{n}_{p}}
\end{equation}
where $r_{p}^{~} \approx 1.53\times 10^{-18}$~m is the classical radius of the
proton and $\sigma^{~}_{p} \approx \sigma^{~}_{e}$ was used. Since electron
bunches are discarded after each collision, only the tune shift of the proton
beam, $\Delta Q_{p}^{~}$, is considered here. The tune shift is approximately
given by
\begin{equation}
\Delta Q_{p}^{~} \approx 1.2\times 10^{-3}\cdot\frac{N_{e}[10^{10}]}{\mbox
{\large{$\varepsilon$}}^{n}_{p}[10^{-6}\,m]}
\end{equation}
The parameter $\Delta Q_{p}^{~}$ must be limited to about $4 \times 10^{-3}$
in order to stem the emittance growth due to random fluctuations of the electron
bunch parameters \cite{brinkmann}. This imposes an upper limit of $N_{e} \lsim 3
\times 10^{10}$ if one assumes $\mbox{\large{$\varepsilon$}}^{n}_{p} \approx
10^{-6}$~m (see also Table 4 in \cite{yavas}).

\begin{table}[!t]
\centering
\ttabbox{\caption{Parameters of the proposed linac-ring $ep$
collider.}\label{tab:table3}}
   {\begin{tabular}{lllr}
\noalign{\vskip 1mm}   
\hline
\hline
\noalign{\vskip 2mm}
\multicolumn{4}{c} {Electron beam parameters} \\
\noalign{\vskip 2mm}
\hline
\noalign{\vskip 1mm}
Beam energy & E$_{e}$ & GeV & 500\\
\noalign{\vskip 1mm}
\hline
\noalign{\vskip 1mm}
Initial bunch population & $N_{e}$ & $\times 10^{10}$ & 2\\
\noalign{\vskip 1mm}
\hline
\noalign{\vskip 1mm}
Number of bunches & $N_{b,e}$ & ~ & 3200\\
\noalign{\vskip 1mm}
\hline
\noalign{\vskip 1mm}
Bunch interval & $\Delta t_{b,e}$ & ns & 211.376\\
\noalign{\vskip 1mm}
\hline
\noalign{\vskip 1mm}
RF frequency & $f_{\mbox{\tiny{RF}}}^{~}$ & MHz & 1301\\
\noalign{\vskip 1mm}
\hline
\noalign{\vskip 1mm}
Pulse repetition rate & $f_{\rm rep}^{~}$ & Hz & 5\\
\noalign{\vskip 1mm}
\hline
\noalign{\vskip 1mm}
Duty cycle & $d$ & \% & 0.34\\
\noalign{\vskip 1mm}
\hline
\noalign{\vskip 1mm}
Beam power & ${\cal P}_{e}^{~}$ & MW & 25.5\\
\hline
\noalign{\vskip 2mm}
\multicolumn{4}{c} {Proton beam parameters} \\
\noalign{\vskip 2mm}
\hline
\noalign{\vskip 1mm}
Beam energy & E$_{p}$ & TeV & 50\\
\noalign{\vskip 1mm}
\hline
\noalign{\vskip 1mm}
Initial bunch population & $N_{p}$ & $\times 10^{10}$ & 10\\
\noalign{\vskip 1mm}
\hline
\noalign{\vskip 1mm}
Number of bunches & $N_{b,p}$ & ~ & 5300\\
\noalign{\vskip 1mm}
\hline
\noalign{\vskip 1mm}
RMS bunch length & $\sigma_{z,p}^{~}$ & mm & 80\\
\noalign{\vskip 1mm}
\hline
\noalign{\vskip 1mm}
Bunch interval & $\Delta t_{b,p}^{~}$ & ns & 49.7355\\
\noalign{\vskip 1mm}
\hline
\noalign{\vskip 1mm}
RF frequency & $f_{\mbox{\tiny{RF}}}^{~}$ & MHz & 401.968\\
\noalign{\vskip 1mm}
\hline
\noalign{\vskip 2mm}
\multicolumn{4}{c} {Collider parameters} \\
\noalign{\vskip 2mm}
\hline
\noalign{\vskip 1mm}
Beta function at IP & $\beta^{*}_{p}$ & m & 0.1\\
\noalign{\vskip 1mm}
\hline
\noalign{\vskip 1mm}
Norm. transverse emittance & $\mbox{\large{$\varepsilon$}}^{n}_{p}$ & $\mu$m & 1\\
\noalign{\vskip 1mm}
\hline
\noalign{\vskip 1mm}
Beam-beam tune shift & $\Delta Q_{p}^{~}$ & ~ & 0.0024\\
\noalign{\vskip 1mm}
\hline
\noalign{\vskip 1mm}
Electron beam disruption & $D_{e}^{~}$ & ~ & 11.3\\
\noalign{\vskip 1mm}
\hline
\noalign{\vskip 1mm}
Hourglass factor & $H_{\rm hourglass}$ & ~ & 0.81\\
\noalign{\vskip 1mm}
\hline
\noalign{\vskip 1mm}
Pinch factor & $H_{\rm pinch}$ & ~ & 1.3\\
\noalign{\vskip 1mm}
\hline
\noalign{\vskip 1mm}
Proton filling & $H_{\rm filling}$ & ~ & 0.79\\
\noalign{\vskip 1mm}
\hline
\noalign{\vskip 1mm}
Luminosity & ${\cal L}_{ep}$ & $\times 10^{32}~cm^{-2}\,s^{-1}$ & 1.08\\
\noalign{\vskip 1mm}
\hline
\hline
   \end{tabular}} 
\end{table}

A small error $\Delta k$ in the quadrupole gradient leads to a tune shift
$\Delta Q_{k}$. To a beam particle with momentum $p = p_{0}^{~} + \Delta p$ it
appears that all the quadrupoles in the ring have a quadrupole error proportional
to $\Delta p/p_{0}^{~}$ \cite{wille}. The dimensionless quantity $\xi$ defined by
$\Delta Q_{k} \equiv \xi(\Delta p/p_{0}^{~})$ is called the {\em chromaticity} of
the beam optics. This quantity increases with the strength of the beam focusing.
The main contribution to the chromaticity comes from the final focus quadrupoles,
where the $\beta$-function is large \cite{zimmermann5}:
\begin{equation}
\xi \approx \beta_{q}^{~}k_{q}^{~}\ell_{q}^{~} \approx \frac{\ell^{*} + \ell_{q} ^{~}/2}{\beta_{y}^{*}}
\end{equation}
Here $\beta_{q}^{~}$, $k_{q}^{~}$ and $\ell_{q}^{~}$ denote the beta function,
field gradient and length of the final quadrupole, respectively; $\ell^{*}$ is
the focal length and $\beta_{y}^{*}$ the value of the vertical $\beta$-function at
the interaction point. Thus, the chromaticity increases as $\beta_{y}^{*}$
decreases.

Since $\xi$ grows linearly with the distance between the final-focus quadrupole
and the interaction point, it is desirable to make this distance as small as
possible. For the interaction region at an electron-proton collider, a novel
design technique called the {\em achromatic telescopic squeezing} (ATS) has been
proposed ``in order to find the optimal solution that would produce the highest
luminosity while controlling the chromaticity, minimizing the synchrotron
radiation power and maintaining the dynamic aperture required for [beam]
stability" \cite{fartoukh}\cite{cruz-alaniz} ({\em dynamic aperture} is the stability
region of phase space in a circular accelerator).

The issue of beam stability was addressed earlier concerning the optimization of the
proton bunch length. The proton bunches inside the ILC-type linac shown
in Figs 2 and 3 are much shorter than those inside the FCC storage ring (the 3-GeV
injector linac mentioned earlier would deliver bunches a few millimetres long).
Thus, $\sigma_{z,p}$ has to be increased in order to attain the baseline FCC-pp
value (see Table 2). In principle, the easiest way to increase the bunch length in
a circular accelerator is to switch all RF systems off and let the bunches `decay'
due to {\em dispersion}.\footnote{A particle with a momentum difference $\Delta
p/p$ has a transverse position $x(s) + D(s)\Delta p/p$, where $x(s)$ is the
position a particle of nominal momentum would have and $D(s)$ is the
{\em dispersion function}.} A faster and more subtle method --- which could be
implemented using a 3-TeV proton booster placed inside the FCC tunnel --- is
described in \cite{damerau}.

The analytic expressions for beam-beam tune shift, electron beam disruption and
beam growth rate given above do not accurately describe the {\em time-dependent
beam dynamics} during collisions. To study the time-dependent effects caused by
varying beam sizes, collision point simulations for linac-ring $ep$ colliders have
been performed using the ALOHEP software \cite{alohep}. This numerical program
optimizes a set of electron and proton beam parameters in order to maximize
luminosity. Some of the results obtained by the program are presented in
\cite{acar}.

The luminosity ${\cal L}_{ep}$ is independent of the electron bunch charge and
the collision frequency as long as their product, expressed in terms of the beam
power ${\cal P}_{e}$, is constant. One can therefore rewrite Eq. (2) as
follows \cite{tigner}\cite{THERA}
\begin{equation}
{\cal L}_{ep} = 4.8\times 10^{30}\,{\rm cm}^{-2}\,{\rm s}^{-1}\cdot\frac{N_{p}}
{10^{11}}\frac{10^{-6}\,{\rm m}}{\mbox{\large{$\varepsilon$}}^{n}_{p}}\frac
{\gamma_{p}^{~}}{1066}\frac{10\,{\rm cm}}{\beta_{p}^{*}}\frac{{\cal P}_{e}}
{22.6\,{\rm MW}}\frac{250\,{\rm GeV}}{{\rm E}_{e}}\cdot{\cal H}
\end{equation}
The electron beam current $I_{e} = {\rm e}N_{e}f_{b,e} = 15$~mA, where $f_{b,e}$
is the inverse of the bunch interval (see Table 3). The electron beam power
${\cal P}_{e}^{~} ={\rm E}_{e}I_{e}d = 25.5$~MW, where $d$ is the linac duty cycle.
The proton beam current $I_{p} = 320$~mA, and the total energy stored per
proton beam is 4.2 GJ. To calculate $H_{\rm hourglass}$, we set
$\beta^{*}_{e} \approx \beta^ {*}_{p}$ \cite{cruz-alaniz}. The value of
$H_{\rm pinch}$  was taken from \cite{bruning}.

\vspace*{0.3cm}
\section{Multi-MW proton beams for fixed-target experiments at FCC}
\vspace*{0.3cm}

~~~~To search for new physics beyond the Standard Model usually requires the use
of high-energy hadron or electron-positron colliders. However, many important
discoveries in particle physics have been made using proton beams with relatively
low energies but high intensities. With this in mind, it is envisaged that the
facilities described in this note could also provide multi-MW proton beams for
fixed-target experiments to study neutrino oscillations, rare kaon decays, the
electric and magnetic dipole moments of the muon, etc.

The beam power in a circular accelerator is limited by space charge effects that
produce beam instabilities. This limitation is intrinsic to the existing proton
synchrotron complexes at CERN, Fermilab and J-PARC. To increase maximally the beam
power of a `proton driver', one could use the 3-GeV injector linac and the
high-energy {\em pulsed} superconducting L-band structures shown in Figures
\ref{fig:SLC} and \ref{fig:2linac}. A similar facility was originally proposed at KEK
by the present author \cite{belusevic1}.

For a pulsed linear accelerator, the following expression holds:
\begin{equation}
{\cal P}_{\rm b}^{~}\,\mbox{[MW]} = {\rm E}_{\rm b}^{~}\,\mbox{[MV]}\times
I\,\mbox{[A]}\times \tau_{\rm p}^{~}\,\mbox{[s]} \times {\cal R}\,\mbox{[Hz]}
\end{equation}
where ${\cal P}_{\rm b}^{~}$ is the {\em beam power}, ${\rm E}_{\rm b}$ is
the {\em beam energy}, $I$ is the {\em average current per pulse}, $\tau_{\rm p}
^{~}$ is the {\em beam pulse length}, and ${\cal R}$ is the {\em pulse repetition
rate}. Assuming ${\rm E}_{\rm b}^{~} = 60\times 10^{3}$ MV, $I = 25$ mA,
$\tau_{\rm p}^{~} = 1.2$ ms and ${\cal R} = 10~{\rm s}^{-1}$, expression (14)
yields  ${\cal P}_{\rm b}^{~} = 18$ MW. This is more than an order of magnitude
higher than the beam power at any existing proton synchrotron complex. The physics
potential of a multi-MW proton driver is discussed, for instance, in the references
cited in \cite{belusevic1}.

\vspace*{0.5cm}
\section{Acknowledgements}
\vspace*{0.3cm}

~~~~I am grateful to K. Oide and D. Zhou for helping me estimate some relevant
beam properties at the SLC-type facility described in this note, and would like to
thank K. Yokoya for his valuable comments and suggestions.

\end{document}